\documentclass[letterpaper, 10 pt, conference]{ieeeconf}  

\IEEEoverridecommandlockouts                              

\usepackage{graphicx}
\usepackage{gensymb}
\usepackage{amsmath}
\usepackage{cite}
\title{\LARGE \bf
``It was Tragic'': Exploring the Impact of a Robot's Shutdown}

\author{Agam Oberlender and Hadas Erel
\thanks{*This work was not supported by any organization}
\thanks{Media Innovation lab, Reichman University, Herzliya. {\tt\small agam.oberlender@milab.runi.ac.il; hadas.erel@milab.runi.ac.il}}%
}

\begin{document}

\maketitle
\thispagestyle{empty}
\pagestyle{empty}


\begin{abstract}
It is well established that people perceive robots as social entities, even when they are not designed for social interaction. We evaluated whether the social interpretation of robotic gestures should also be considered when turning off a robot. In the experiment, participants engaged in a brief preliminary neutral interaction while a robotic arm showed interest in their actions. At the end of the task, participants were asked to turn off the robotic arm under two conditions: (1) a \textit{Non-designed} condition, where all of the robot's engines were immediately and simultaneously turned off, as robots typically shut down; (2) a \textit{Designed} condition, where the robot’s engines gradually folded inward in a motion resembling ``falling asleep." Our findings revealed that all participants anthropomorphized the robot's movement when it was turned off. In the \textit{Non-designed} condition, most participants interpreted the robot’s turn-off movement negatively as if the robot had ``died." In the \textit{Designed} condition, most of the participants interpreted it more neutrally, stating that the robot: ``went to sleep." The robot's turn-off movement also impacted its perception, leading to higher likeability, perceived intelligence, and animacy in the \textit{Designed} condition. We conclude that the impact of common edge interactions, such as turning off a robot, should be carefully designed while considering people's automatic tendency to perceive robots as social entities.

\end{abstract}

\section{Introduction}

As robots become increasingly integrated into daily life, it is essential to consider how people perceive and engage with them socially. Although robots are not necessarily designed for social interactions, research indicates that humans tend to perceive them as social entities \cite{hoffman2014designing, erel2019robots, erel2021excluded, erel2024rosi, novikova2014design}. People automatically interpret robots' movements as social cues \cite{erel2019robots, hoffman2014designing} even when the robot is highly simple and abstract in form \cite{hamacher2016believing, fukuda2012midas, iio2011investigating, saunderson2019robots, adi2022non, jung2020robot, hoffman2014designing}. These studies indicate that interactions with non-humanoid robots involve more than just functional aspects and that their mechanical attributes are not the only factors shaping their perception. Participants often report significant social engagement with non-humanoid robots, despite acknowledging their machine-like nature \cite{erel2022carryover, erel2021excluded, zuckerman2020companionship}. For example, gestures performed by robots designed as a desk lamp \cite{rifinski2021human}, a ball rolling on a dome \cite{anderson2018greeting}, or a robotic arm \cite{jung2020robot} have been interpreted as indicating care \cite{zuckerman2020companionship}, attentiveness \cite{rifinski2021human}, interest \cite{manor2022non, tennent2019micbot}, and a general willingness to interact with the participant \cite{anderson2018greeting}. The strong association between robots' movements and the social experience when interacting with them is typically explained by the human tendency to perceive the world through a social lens and anthropomorphize objects in their environment \cite{erel2019robots}. A robot's autonomous behavior strengthens this tendency, leading to the perception of movements as social cues reflecting the robot's intentions and general inner state \cite{erel2019robots, gazzola2007anthropomorphic, dunbar1998social, duffy2003anthropomorphism}.


 \begin{figure}[t]
 \vspace{0.2cm}
\includegraphics[width=1\linewidth]{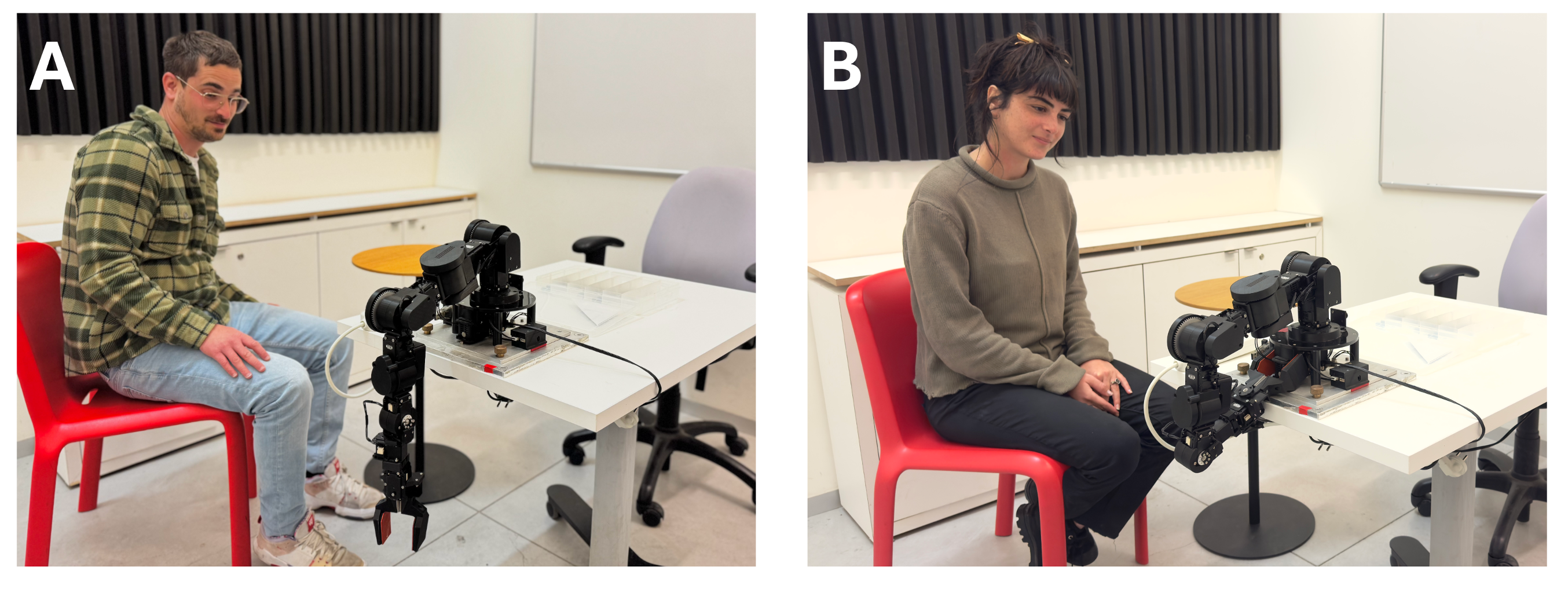}
\caption{(A) Participant experiencing a standard robot turn-off ; (B) Participant experiencing a designed robot turn-off}
     \label{condition}
    \vspace{-1em}
 \end{figure}

Robots' anthropomorphism has many advantages, and it is already leveraged for designing robotic behaviors that would be easily understood and accepted \cite{hoffman2014designing, duffy2003anthropomorphism, erel2022carryover, kumaran2023cross, chakravarthi2024social}. At the same time, it presents several challenges that should be carefully addressed. Among these challenges is the need to design robotic behaviors that comply with social norms and to consider the social interpretation of the robot's gestures, even when its purpose does not involve any social aspects \cite{tennent2019micbot, erel2024power, gillet2024interaction}. Failing to account for such social perspectives can lead to intense adverse effects that would shape the robot's perception \cite{paetzel2020persistence, avelino2021break, anderson2018greeting, heimann2025scammed, erel2024power, manor2024attentiveness} and impact the humans that interact with it \cite{erel2021excluded, manor2022non, hitron2022ai, hitron2023implications, yaar2025s}. These adverse effects may also extend to future interactions with robots or with other humans \cite{erel2024power, erel2022carryover, heimann2025scammed}. Consequently, when designing robots, it is essential to look beyond the robot's functionality and also take into account the social interpretation of its movements and behavior.

One aspect that is commonly overlooked when designing robots and their behavior is the robot's life cycle \cite{kamino2023towards, laity2024rust, kamino2024lifecycle}. Designers and researchers often focus their efforts on adjusting the robot's appearance and movement to support its purpose and primary function \cite{hoffman2014designing}. Robot edge cases, ranging from simple malfunctions to turning off the robot and the robot's ``end-of-life", are rarely addressed and included in the design process. The perception of robots as social entities implies that experiences involving such edge cases may result in drastic effects, especially if not carefully designed \cite{kamino2023towards, laity2024rust}. This idea is supported by a handful of studies that explored humans' experiences of robots' malfunctions \cite{kamino2025robot, kamino2024lifecycle} and ``end-of-life" \cite{laity2024rust, fraser2019we, carter2020death, carpenter2013quiet}. When encountering these edge cases, people demonstrated strong emotional responses, such as grief, guilt, anxiety, and moral conflicts \cite{kamino2023towards, carter2020death, ostrowski2022mixed, holland2021my, laity2024rust}. Interestingly, to the best of our knowledge, the direct interpretation and the emotional responses to turning off a robot (or a robot that runs out of battery) have not been explored. Previous studies imply that these (non-designed) edge cases may result in more than simple technical experiences. People commonly hesitate and show discomfort when asked to turn off a robot \cite{bartneck2007daisy, horstmann2018robot, Horsi2024accepting}. Moreover, the time it takes to turn off a robot is considered an indicator of the robot's animacy \cite{bartneck2007daisy} and the participant's empathy toward the robot \cite{hri2019switch}. This suggests that humans tend to add social interpretation and somewhat anthropomorphize the action of turning off a robot, with some researchers describing it as a ``temporary death" \cite{laity2024rust}. It is, therefore, essential to evaluate people's interpretation of the standard robotic turn-off movement, where all engines shut down simultaneously, and its implications for the robot's perception.

In this study, we tested people's experience when turning off a robot. We evaluated whether they anthropomorphize it and attribute social meaning to the robot's movement when turned off. We designed a brief preliminary (neutral) interaction with a robotic arm and then asked participants to turn it off. We compared two conditions: a standard \textit{Non-designed} turn-off, where all engines shut down immediately and simultaneously (as robots typically shut down); and a \textit{Designed} condition, where the robot's gestures were designed to indicate a slow and gradual shutdown, intended to create an experience of the robot ``falling asleep" (see Figure \ref{condition}). We intentionally used an industrial mechanical robotic arm to test whether the social interpretation of the turn-off movement also applies to highly non-humanoid robots. In the experiment, participants were asked to sort screws into groups according to size. Throughout the task, a robotic arm positioned on a table in front of them performed gestures that appeared to follow their actions, as if it was interested in their sorting process. When participants completed the task, they were asked to turn off the robotic arm using a turn-off button. We evaluated participants' experience in the different conditions, their interpretation of the robot's movement when turned off, and their associated perception of the robot. By comparing the two conditions, we tested whether a robot's turn-off movement should be carefully considered and well-designed.

\begin{figure*}[t]
     \centering
\includegraphics[width=0.94\linewidth]{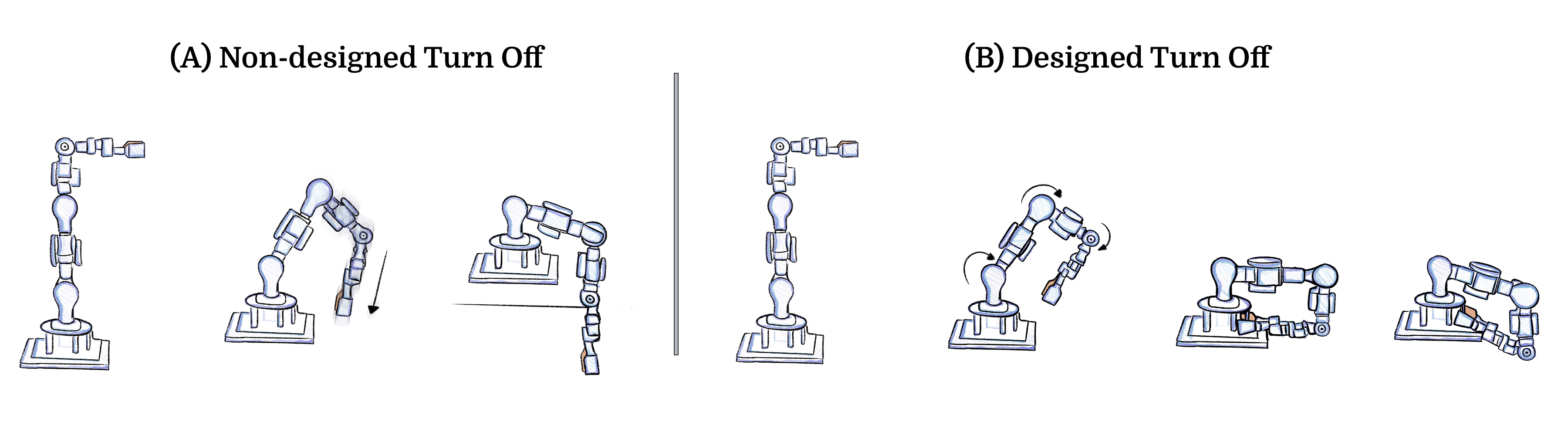}
\caption{The robot's turn-off gestures: (A) \textit{Non-designed} condition; (B) \textit{Designed} condition}
     \label{fig:gestures}
    \vspace{-1em}
 \end{figure*}

\section{Related Work}
Relevant previous work includes studies that involved turning off robots, robotic edge cases, and social interpretation of robots' nonverbal gestures.

\subsection{Turning off robots}
A few studies indicated that turning off a robot is an action that may involve robot anthropomorphism and social experiences \cite{wieringa2023peg, sorengaard2024switching, bartneck2007daisy, horstmann2018robot, hri2019switch, Horsi2024accepting}. For example, Őrsi and colleagues (2024) evaluated how attitudes toward robots affect compliance with a robot's request not to be turned off. Their findings showed that positive attitudes toward robots increased participants' compliance with the robot's request, whereas negative attitudes increased the likelihood of turning it off \cite{Horsi2024accepting}. Similarly, Hostmann et al. (2018) showed that when a robot expressed emotional protest to being turned off, participants tended to attribute more human-like traits to it and kept it turned on \cite{horstmann2018robot}. The robot's characteristics were also indicated as factors influencing participants' willingness to turn off a robot. For example, Bartneck et al. (2007) found that participants were more hesitant to turn off an intelligent and pleasant robot than a less agreeable and less intelligent one \cite{bartneck2007daisy}.

These findings imply that turning off a robot is not simply a technical action and that social aspects, such as the robot's behavior and characteristics, shape people's willingness to turn it off. We extend these studies by revealing the underlying process behind this social experience, characterized by participants’ immediate responses and their dramatically negative interpretation of the robot's standard turn-off action.

\raggedbottom

\subsection{Robotic edge cases} 
Several studies have examined people's responses to robotic edge cases. These studies suggest that people demonstrate intense emotional reactions when encountering different stages of a robot's life cycle, including malfunctions and ``end-of-life" \cite{kamino2023towards, carter2020death, ostrowski2022mixed, holland2021my, laity2024rust, kamino2025robot}. 
A well-known example includes people's broad and intense reactions to the destruction of hitchBOT \cite{fraser2019we}, a speaking but immobile robot that `traveled' by hitchhiking across various countries until it was destroyed by vandals. The analysis of the public reactions to hitchBOT's destruction showed that people experienced intense negative emotions, driven by a strong emotional connection to hitchBOT, and perceived its destruction as morally wrong. This was reflected in the anthropomorphic language used to describe the event, with terms like ``die," ``death," ``demise," ``killing," ``kill," and even ``murdered" \cite{fraser2019we}. Similarly, Carter et al. (2020) tested how people respond on Twitter to the deaths of humans (celebrities), animals (cats), and robots (Mars Opportunity Rover, Jibo, and Kuri) \cite{carter2020death}. Their findings suggested that people often use similar language when reacting to the `deaths' of robots and to the deaths of humans. Most responses to robots' endings were framed using second-person pronouns (``you") and third-person pronouns (``she/he" instead of ``it") \cite{carter2020death}. Another line of studies indicated people's intense emotional responses when asked to destroy a robot. For example, Bartneck et al. (2007) tested how the perceived intelligence of a robot influenced participants' willingness to destroy it. Participants interacted with a `smart' or `stupid' robot and were then asked to destroy it with a hammer. The findings revealed that participants experienced negative emotions, including guilt and shame, when they destroyed the smart robot \cite{bartneck2007kill}. Similarly, Wieringa et al. (2023) showed that participants felt more guilty after destroying a robot in comparison to a non-responsive object \cite{wieringa2023peg}.

Another robotic edge case involves robots' malfunctioning. Kamino and colleagues (2024), interviewed robot owners alongside sales and repair companies and mapped their experiences with robots' lifespan, repair, and obsolescence. Their interviews revealed that robot repair involves more than just restoring its functioning. They suggested that it reflects the owner-robot relationship and their emotional attachment to it \cite{kamino2024lifecycle}. In a later study, the authors found that owners frequently referred to robot repair using medical terms like “treatment” and “hospitalization,” highlighting the robot’s role as a social entity. They also indicated that, as a result, robotic companies are already offering social solutions to robot owners, including constant updates on the robot's repair progress and funeral services if the robot cannot be repaired \cite{kamino2025robot}.


These studies indicate people's intense responses to two major robotic edge cases, a robot's ending and malfunctioning. We extend this research by testing people's interpretation and responses to the more common edge case of turning off a robot and evaluate whether it should be carefully designed.

\subsection{Social interpretation of robots' nonverbal gestures}
Previous studies indicated people's automatic tendency to perceive interactions with non-humanoid robots as social experiences \cite{erel2021excluded, novikova2014design}. These studies showed that robots' non-verbal gestures are commonly perceived as clear and consistent social cues \cite{novikova2014design, erel2022enhancing, jung2020robot, zuckerman2020companionship}.

For example, Erel et al. (2019) showed that the social interpretation of an abstract robot's movements involves an automatic cognitive process that people cannot avoid. In their study, participants were instructed to categorize the robot's gestures (a small ball rolling in a dome) into directional categories, identifying whether the ball's movement was toward the front or back of the dome. Using the Implicit Association Test, they found that participants consistently perceived the robot's movements as social cues, indicating the robot's willingness to interact. This was the case even when the social interpretation of the movement conflicted with the direction classification task participants were asked to perform \cite{erel2019robots}. Similarly, Ju and Takayama (2009) showed that people attribute social interpretation to the movement of an automatic door. When the door opened in response to the participants’ proximity, participants perceived it as a greeting and welcoming gesture \cite{ju2009approachability}. Intense social experiences were also indicated with a lamp-like robot that performed 'lean,' 'gaze,' and 'nod' gestures toward participants sitting in front of it. Participants interpreted the robot's gestures as indicating the robot's attentiveness and care \cite{manor2022non, manor2024attentiveness}.
Another line of studies indicated the negative effects of robots that do not comply with social norms. For example, Erel et al. (2024) showed that when a robot does not perform a greeting gesture at the opening encounter, it decreases the quality of the interaction that follows and participants' willingness to help the robot \cite{erel2024power}. 

These studies show that nonverbal gestures are inherently interpreted as social cues, influencing people's experience and the perception of the robot. We extend these studies by testing whether a robot's movement, when turned off, is anthropomorphized and whether it triggers a social experience, even if the robot has no humanoid features.

\section{Method}

To examine whether people anthropomorphize and attribute social meaning to a robot's movement when turned off, we designed a study that included a brief preliminary (neutral) interaction with a robotic arm and then asked participants to turn off the robot. In the task, participants were instructed to sort screws into groups of similar size. A robotic arm placed on the desk behind the screws `showed interest' in the participants' sorting by performing gestures that followed their sorting movements. As participants finished their task, they were asked to turn off the robotic arm using a turn-off button. Their action either turned off all of the robot's engines simultaneously (as robots typically shut down) -- \textit{Non-designed} condition (see Figure \ref{fig:gestures}A), or triggered a pre-designed gradual movement inward into a converging position, and only then turned off all engines -- \textit{Designed} condition (see Figure \ref{fig:gestures}B). We evaluated participants' experience in the different conditions and their perception of the robot.

\subsection{Robot and Implementation}
We used a Cyton Gamma 1500 robotic arm, with seven degrees of freedom (see Figure \ref{fig:setting}). We intentionally chose an industrial robotic arm to evaluate whether a social interpretation of a robot's turn-off experience should be considered even if the robot has no humanoid features. The specific choice of a robotic arm also allowed us to employ a previously studied gesture that is known to be interpreted as indicating that the robot is ``falling asleep" \cite{ende2011human}, which could serve as an alternative well-designed turn-off process for the standard immediate turn off of all engines. 

\subsubsection{Implementation}
We used the Butter Robotics platform as the robot’s infrastructure \cite{benny2020}. The seven degrees of freedom of the robot were daisy-chained together and terminated in the Butter Robotics hardware controller. The Butter Composer directly translated Blender animations to engine movements, and the robotic arm was controlled wirelessly using a Wizard of Oz (WoZ) technique \cite{riek2012wizard}.

\subsubsection{Gesture Design}
The robot's carefully designed alternative gesture was developed through multiple iterations, and based on previous literature \cite{ende2011human}. An animator and an HRI expert participated in the design process, which focused on body language and sleep-like gestures.
The process resulted in a gradual movement leading to the folding of the entire arm. Each part bends and curves in a slow, smooth, continuous motion until the robotic arm fully folds into itself, appearing to converge inward into a closed state. This gesture sequence was followed by the turning off of all engines (see Figure \ref{fig:gestures}B).

The team also designed gestures for the preliminary interaction that included:
 \begin{itemize}
   
     \item \textit{Greeting}: The robotic arm leaned towards the participant, followed by a double up-down movement of the top part, simulating a nodding. 
     \item \textit{Observe Long}: The robotic arm initially assumed a curved posture, towards the participant. Then, it leaned toward the screws and performed a scanning motion, rotating 30\degree\ to the right and then to the left, repeating this sequence twice to simulate a scanning behavior. This motion lasted approximately 11 seconds, after which the arm returned to its starting position (a semi-curved position toward the participant). 
     \item \textit{Observe Short}: The robotic arm performed a similar movement but rotated 15\degree\ to the right and left, repeating this sequence twice to simulate a scanning behavior. This motion lasted approximately 9 seconds, after which the arm returned to its starting position (a semi-curved position toward the participant).
     \item \textit{Interested}: The robotic arm leaned toward the screws and performed a double up-down movement of the top part from the screws towards the participant and back. This motion lasted approximately 8 seconds, after which the arm returned to its starting position (a semi-curved position toward the participant).
 \end{itemize}

The understanding of all gestures was verified in a pilot study that included 15 participants. Participants watched each gesture and described its meaning. All participants easily understood the gestures and provided consistent interpretations. 
The robot's gestures were sequenced into a fluent robotic behavior using a WoZ technique. In both conditions, the robotic arm performed a \textit{Greeting} gesture as the participants sat next to the table. Then, as participants began the task, the robotic arm consistently performed the following animations in sequence: \textit{Observe Long}, \textit{Interested}, \textit{Observe Short}, and \textit{Interested}, with a 10-second pause between each gesture. Upon task completion, the robotic arm returned to its zero position (upright). The understanding of the gestures was validated using a short interview.

\subsection{Participants}
Forty undergraduate students from the university participated in the study (20 females, 20 males; mean age = 23.5, SD = 1.9). All participants completed an informed consent form and received additional course credits for their participation.

 \begin{figure}[t]
     \centering
      \vspace{0.2cm}
    \includegraphics[width=0.90\linewidth]{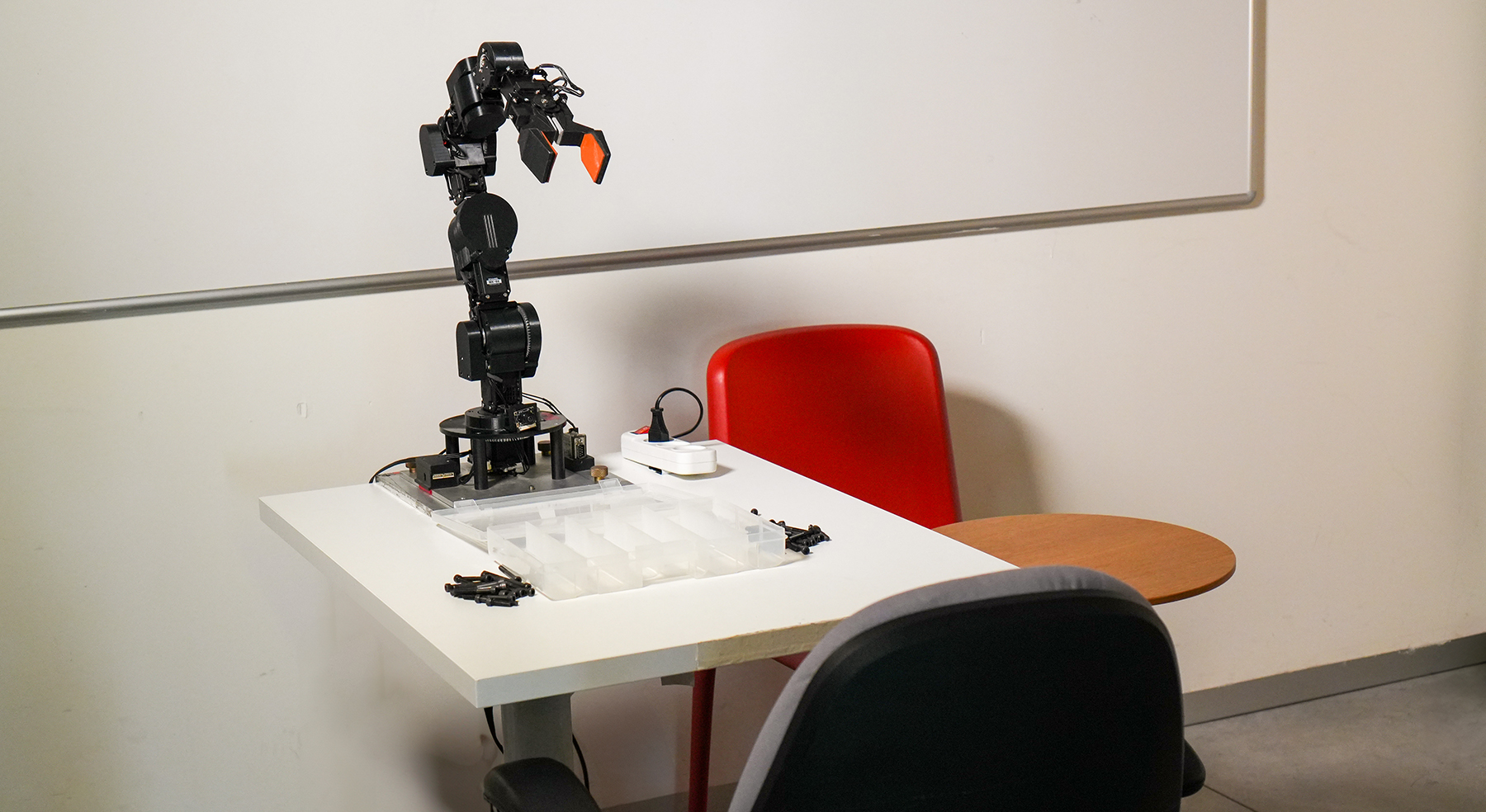}
     \caption{Experimental settings}
     \label{fig:setting}
    \vspace{-1em}
 \end{figure}

\subsection{Experimental design}
A between-participant experimental design included two similar conditions apart from the turn-off movement (see Figure \ref{condition}). In the \textit{Non-designed} condition, when the participant pressed the turn-off button, all the robot's engines were immediately turned off (as robots typically shut down; see Figure \ref{fig:gestures}A). In the \textit{Designed} condition, the robot first performed the \textit{Designed} turn-off movement where the engines gradually moved inward into a converging position and then the engines turned off (see Figure \ref{fig:gestures}B). To avoid a priori differences between groups, participants were \textit{randomly} assigned to conditions using a matching technique that balanced gender and pre-test scores of the Negative Attitudes Toward Robots (NARS) questionnaire \cite{nomura2006experimental}.

\subsection{Experimental settings}
The experiment took place in a quiet room (see Figure \ref{fig:setting}). A 60 × 80 cm white table was positioned at the center of the room. The robotic arm was placed on the table approximately 55 cm vertically from the participant’s head, with a power switch located on the right of the robotic arm. A plastic box with screws on its sides was positioned between the participant and the robotic arm. On the right side of the room, another chair and a small table were set up for the second part of the experiment, where participants completed questionnaires. Additionally, a small camera was placed on a table at the back of the room.

\subsection{Measures}
Quantitative and qualitative measures were used to assess participants' experience and perception of the robotic arm after turning it off.

\subsubsection{Spontaneous mentions of negative and neutral aspects of the turn-off movement}
To evaluate how participants perceived the robot’s turn-off movement, we coded the frequency of participants who spontaneously anthropomorphized the turn-off movement using either negative or neutral terms when asked to describe their experience.

\subsubsection {Godspeed questionnaire}
In order to test participants' perception of the robot we used the Godspeed questionnaire \cite{bartneck2009measurement}, a widely used semantic differential scale in HRI studies. We used the likability (Cronbach's alpha = 0.81), animacy (Cronbach's alpha = 0.79), and perceived intelligence (Cronbach's alpha = 0.85) subscales.

\subsubsection{Semi-structured interviews}
We conducted a semi-structured interview in order to gain a deeper understanding of participants' emotions, interpretations, and attitudes during the interaction \cite{boyatzis1998transforming, galletta2013mastering}. The interviews were conducted according to clear guidelines (see \cite{knott2022interviews}). The questions included were: ``Describe your experience," ``Describe your thoughts about the robotic arm," ``Describe what happened when you were asked to turn-off the robot," and ``How would you describe this experiment to a friend?"

\subsection{Procedure}

A few days before the experiment, participants received the NARS questionnaire \cite{nomura2006experimental} via email. When participants arrived at the lab, they were informed that the session would be recorded from that point onward and that they could withdraw at any time without penalty. They were then asked to complete a consent form and a demographic questionnaire. Following this, participants were invited into the experiment room and seated facing the screws, the sorting box, and the robotic arm, which then performed the \textit{Greeting} gesture. The researcher explained that they were asked to sort screws of different sizes into designated areas within the plastic box placed between them and the robotic arm. Long screws were to be placed on the right side of the box, while short screws were placed on the left. The researcher also explained that the robotic arm was being trained on sorting and that it collected data as part of its training. As participants began the task, the robotic arm consistently performed the pre-designed animation sequence (\textit{Observe Long}, \textit{Interested}, \textit{Observe Short}, and \textit{Interested}, with a 10-second pause between each gesture). The interaction lasted approximately two minutes (mean = 2.3, SD = 0.5). 

Upon completing the sorting task, the researcher returned to the room and asked the participant to move to the table on the right to fill out questionnaires. The researcher then asked the participant (who now sat next to the turn-off button) to turn off the robotic arm by pressing the power switch to ``prevent overheating." The robotic arm then performed the relevant turn-off behavior based on the relevant condition (\textit{Non-designed} or \textit{Designed}). Following the turn-off phase, participants completed the Godspeed questionnaire \cite{bartneck2009measurement}, and participated in the semi-structured interview. At the end of the experiment, participants were invited to share a recent positive experience (as a way to mitigate any potential negative effects) and the researcher debriefed the participants.

\section{Analysis}
To verify the lack of early differences between groups, we conducted Bayesian analyses on the results of the pre-test. Our primary analysis tested participants' perception of the robot's turn-off movement. We conducted a chi-square test to analyze the frequency of participants who used negative versus neutral terminology when spontaneously discussing how the robot turned off (after simply being asked to describe their experience). To evaluate the participants' perception of the robotic arm, we conducted independent t-test analyses for the three Godspeed subscales \cite{bartneck2009measurement}.

\section{Findings}
The Bayesian analysis indicated no early differences between groups on the NARS questionnaire: $BF_{10}=0.2$. The quantitative and qualitative primary analyses revealed that participants anthropomorphized and assigned social interpretation to the robot's turn-off movement.

\begin{table}[b]
\caption{Distribution of participants' use of negative or neutral terms in the different conditions.}
\label{tab:my-table}
\centering
\renewcommand{\arraystretch}{1.5}
\setlength{\tabcolsep}{12pt}
\begin{tabular}{|c|c|c|c|}
\hline
\multicolumn{1}{|c|}{} & \multicolumn{2}{c|}{Social interpretation} & \multicolumn{1}{c|}{} \\ \cline{2-3}
Turn-off condition & Negative & Neutral & Total \\ \hline
Designed & 3 & 17 & 20 \\
Non-designed & 17 & 3 & 20 \\
Total & 20 & 20 & 40 \\ \hline
\end{tabular}
\end{table}

\begin{figure*}[t]
     \centering
\includegraphics[width=\textwidth]{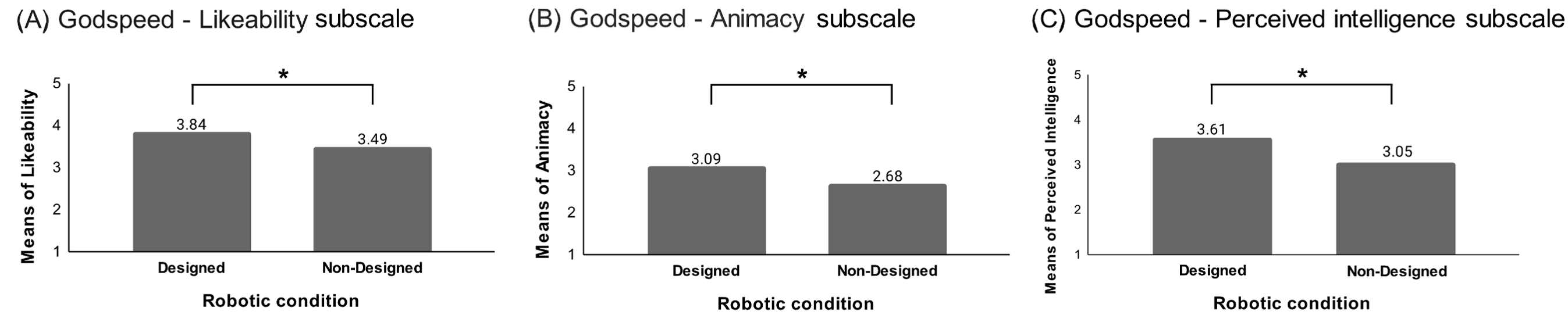}
     \caption{Analysis of the robots' perception; The Godspeed subscales: (A) Likeability, (B) Animacy, (C) Perceived intelligence.}
     \label{fig.Graph}
     \vspace{-1em}
 \end{figure*}

\subsection{Quantitative analysis}

\subsubsection{Social interpretation of the turn-off movement} 
The analysis revealed that the design of the robot's turn-off movement determined the valence of the experience and its social interpretation (see Table \ref{tab:my-table}). While all participants anthropomorphized the movement, their descriptions revealed a great difference in their interpretation valence across the two conditions ${\chi}^2$$_\text{(1)}=19.6, p<0.001$. Most of the participants in the \textit{Non-designed} condition used negative valence terminology (e.g., ``lifeless," ``dead," ``dying") when describing their immediate interpretation of the robot's turn-off movement. In contrast, most of the participants in the \textit{Designed} condition used neutral terminology (e.g., ``resting," ``going to sleep," ``closing itself").

\subsubsection{Robot Perception}

The analyses of the Godspeed subscales revealed that the design of the turn-off movement impacted all three subscales (see Figure \ref{fig.Graph}). In the likeability subscale, $\textit{t}_{(38)} = 1.98, p=0.05$, Cohen's D = 0.63, participants rated the robot as more likable in the \textit{Designed} condition (M = 3.84, SD = 0.64) than the \textit{Non-designed} condition (M = 3.49, SD = 0.45) (see Figure \ref{fig.Graph}A). Similarly in the \textit{animacy} subscale $\textit{t}_{(38)} = 2.46, p<0.01$, Cohen's D = 0.78, participants in the \textit{Designed} condition rated the robot as more animate (M = 3.09, SD = 0.50) than those in the \textit{Non-designed} condition (M = 2.68, SD = 0.54) (see Figure \ref{fig.Graph}B). In the \textit{perceived intelligence} subscale, $\textit{t}_{(38)} = 2.55, p<0.01$, Cohen's D = 0.81, participants in the \textit{Designed} condition rating the robot as more intelligent (M = 3.61, SD = 0.63) than those in the \textit{Non-designed} condition (M = 3.05, SD = 0.74) (see Figure \ref{fig.Graph}C).

\subsection{Qualitative analysis - Semi-structured interviews}
The qualitative analysis of the interviews was conducted using thematic coding methodology \cite{boyatzis1998transforming}. First, two coders transcribed and reviewed the interviews to develop a preliminary understanding of the data. Then, initial themes were identified, and any inconsistencies were resolved through a discussion with a third researcher. The coders independently analyzed part of the data to verify inter-rater reliability (kappa = 0.83\%) and then analyzed the remaining data. 
The thematic analysis revealed three themes: \textit{perception of the robot's turn-off movements}, \textit{responses to the robot's turn-off movements}, and \textit{the need to redesign the robot's movement}. 

\subsubsection {Perception of the robot's turn-off movement}
All participants anthropomorphized the robot’s turn-off movement. However, the valence of their interpretations varied between conditions. Most participants in the \textit{Non-designed} condition (17/20) interpreted the movement negatively, describing it with terms related to human death: \textit{``It looked like it was dying."} (p. 36, M); \textit{``It was about to die."} (p. 21, F). Only a minority of the participants (3/20) perceived the movement more neutrally: \textit{``It was going to sleep."} (p. 18, F).

In the \textit{Designed} condition, most of the participants interpreted the robot's turning off movement more neutrally, describing it as a controlled, intended gesture indicating that the robot is going to sleep or moving to a base position (17/20): \textit{``It was going to sleep. It shut down slowly, closing into itself."} (p. 31, M); \textit{``It gently returned to its place."} (p. 48, F). 3/20 participants perceived the movement negatively: \textit{``I thought it broke down for a moment."} (p. 25, F).

\subsubsection {Responses to the robot's turn-off movements}
Participants responded differently to the robot's turn-off movement. In the \textit{Non-designed} condition, 14/20 participants discussed the turn-off behavior. Most of them (11/14) described a negative response: \textit{``It was tragic. It was kind of sad."} (p. 21, F); \textit{``It felt strange... it was a bit aggressive. The robot looked helpless; something was unsettling about the way it collapsed."} (p. 50, F). Only 3/14 participants responded positively: \textit{``It was interesting and cool."} (p. 40, M). 

In the \textit{Designed} condition, 14/20 participants discussed the turn-off movement. Most participants described positive emotions (10/14): \textit{``It was cool; it calmed me down."} (p. 39, M); \textit{"It's cool it returned to its home area."} (p. 33, M). 4/14 participants responded negatively: \textit{``It was strange, it was there with you, and then it’s not there anymore."} (p. 43, F).

\subsubsection {The need to redesign the robot's movement} 
Several participants expressed a need to redesign the robot's movement when turned off. The overwhelming majority of these participants were from the \textit{Non-designed} condition (18/20). Their spontaneous suggestions were surprisingly similar to the robot's gesture in the \textit{Designed} condition: \textit{``Instead... it could sort of lower itself slowly."} (p .32, M); \textit{``I would make it fold into itself, in a small square-like form."} (p. 16, F). In the \textit{Designed} condition, only 4/20 participants discussed the need to redesign the movement or thought the robot should not move at all: \textit{``I expected it to stay static."} (P. 23, F).

\section{Discussion}
In this work, we demonstrate that people anthropomorphize and attribute social meaning to a robot's movement when it is turned off. This effect occurs even with an industrial robot lacking humanoid features. In both conditions, participants consistently anthropomorphized the robot, reinforcing the notion that even non-humanoid robots are perceived as social entities \cite{hoffman2014designing, erel2019robots, erel2021excluded, erel2024rosi, novikova2014design}. The design of the robot’s turn-off gesture determined the valence of participants' experience in the interaction. In the \textit{Non-designed} condition, most participants described the movement using negative valence terms related to human death, such as ``died," ``lifeless," ``dying," and ``broke its neck." In contrast, the \textit{Designed} turned-off movements were described using more neutral terms such as ``went to sleep" and ``elegantly turned off," indicating that participants perceived the action as controlled and intentional but not negative. We therefore extend the understanding of adverse effects associated with people's tendency to anthropomorphize robots. We indicate that the specific anthropomorphism of turn-off gestures can lead to highly negative experiences directly related to the robot's death. We also provide an alternative and suggest guidelines for the design of movement that would lead to a positive rather than a negative experience in cases where people are required to turn off a robot.

Our findings further suggest that the negative interpretation of the \textit{Non-designed} turn-off movement resulted in negative emotions and created a highly unpleasant experience, with participants describing it as ``dramatic," ``tragic," and even ``aggressive." This highlights that simply turning off all engines immediately and simultaneously can have severe implications, resulting in adverse experiences likely to shape people's emotions. However, if well-designed, a robot's turn-off gestures can lead to a neutral or even pleasant experience, described as  ``calming," ``interesting," and ``cool." The robot's turn-off movement also impacted its perception, and a well-designed movement led to significantly higher ratings of the robot's likability, animacy, and intelligence. 

Our findings align with previous research suggesting that turning off a robot is not simply a technical action \cite{wieringa2023peg, sorengaard2024switching, bartneck2007daisy, horstmann2018robot, hri2019switch, Horsi2024accepting}. We highlight the importance of considering and carefully designing turn-off movements as an integral part of the robot’s behavior, regardless of its function or humanoid appearance. This was also reflected in participants' spontaneous suggestions for design improvements. Participants' ideas closely resembled the gradual turn-off gesture used in the \textit{Designed} condition (which they did not encounter). These consistent descriptions of a more ``appropriate" turn-off movement imply that people may already have clear expectations for robotic behavior in such edge cases and that these should be accommodated when designing robots.

More broadly, we join the call to generally map and design robotic edge cases and the entire robotic life cycle \cite{kamino2023towards, kamino2024lifecycle, kamino2025robot, laity2024rust}. Like robotic breakdowns or ``end-of-life," turning off a robot cannot be overlooked when designing human-robot interactions. The perception of robots as social entities requires a holistic consideration of robots' behavior that extends beyond the specific design of the robot's functionality.

\section{Limitations}

We evaluated participants' responses immediately after the interaction, and they had no opportunity to regulate their emotions. Future studies should evaluate the extent of the effect over time and participants' reactions after multiple turn-off experiences. In addition, participants’ feedback in the interview may have been influenced by the ``good subject effect" \cite{nichols2008good}. We followed a strict protocol to minimize this bias and explained that all answers were valuable. Lastly, while the highly social interpretation of the industrial robot's turn-off gesture implies that such effects would be strengthened with more humanoid robots, the effect should be further tested with different robots in additional contexts.

\section{Conclusion}
We highlight the importance of considering and designing the common edge case of turning off a robot. Like other aspects of the interaction, the robot's movement when turning it off is anthropomorphized. If not well-designed, the turn-off experience is likely to be associated with highly negative interpretations related to death. Our findings call for the careful mapping of robots' edge cases and the explicit design of the robot's behavior while considering the automatic human tendency to perceive robots as social entities.


\addtolength{\textheight}{-5pt}   





\section{Acknowledgements}
A special thanks to Nevo Heimann Saadon, Tomer Etzion, Hila Zohar, Zohar Fein, Ziv Keidar, and Yuval Rubin.

\bibliography{BIBMAIN}
\bibliographystyle{IEEEtran}

\end{document}